\documentclass[12pt,a4paper]{article}
\usepackage{latexsym,amsmath,amssymb}
\date{}
\numberwithin{equation}{section}

\begin{document}
\title{Supersymmetric Gupta-Bleuler Quantisation}
\author{Florin Constantinescu\\ Fachbereich Mathematik \\ Johann Wolfgang Goethe-Universit\"at Frankfurt\\ Robert-Mayer-Strasse 10\\ D 60054
Frankfurt am Main, Germany}
\maketitle

\begin{abstract}
We formulate the Gupta-Bleuler canonical, covariant and gauge invariant supersymmetric quantization of the supersymmetric (massless) vector field. Our main tool is the recently introduced Hilbert-Krein structure of the $N=1$ superspace. 
\end{abstract}

\section{Introduction}

The well known covariant Gupta-Bleuler quantization in quantum electrodynamics (see for instance \cite{IZ}) makes use of an indefinite metric. In its rigorous version \cite{GW, StroW, Stro} it is formulated in the frame of a Hilbert-Krein structure (see also \cite{G}). Although in the modern approach the quantization of gauge theories (abelian and non-abelian) heavily uses functional integration techniques, the Gupta-Bleuler covariant and gauge invariant quantization in the abelian case remains a remarkable achievement of canonical quantization.\\
In this note we extend the Gupta-Bleuler quantization to the supersymmetric case. The nice point of our construction is the appearance of a natural supersymmetric Krein structure which produces by standard methods the physical supersymmetric Hilbert space. Besides the standard text-book knowledge (\cite{IZ,WB} are most adequate for our needs) our tools rely on a simple superspace finding which will be shortly presented below (for more details the reader may consult \cite{C}). As supersymmetries are concerned, we use the conventions and notations of \cite{Sr}. \\
In the second section we shortly review the main points of the supersymmetric Hilbert-Krein structure whereas in the third section we perform the supersymmetric Gupta-Bleuler quantization. \\

\section{The supersymmetric Hilbert-Krein structure}

Let 

\begin{gather} \nonumber
X(z)=X(x,\theta ,\bar \theta )= \\ \nonumber
=f(x)+\theta \varphi (x) +\bar \theta \bar \chi (x) +\theta ^2m(x)+\bar \theta^2n(x)+ \\
\theta \sigma^l\bar \theta v_l(x)+\theta^2\bar \theta \bar \lambda(x)+\bar \theta^2\theta \psi (x)+ \theta^2 \bar \theta^2d(x)
\end{gather}
be a general supersymmetric function with regular coefficients in front of the $ \theta ,\bar \theta $-powers (for instance contained in the Schwartz space $S$) and $ D_\alpha ,\bar D_{\dot \alpha }$ the covariant derivatives in the $N=1$ superspace.
Let
\[c=\bar D^2D^2, a=D^2\bar D^2, T=D^{\alpha }\bar D^2 D_{\alpha }=-8\square +\frac {1}{2}(c+a) \]
and let
\[ P_c=\frac {1}{16\square }c,P_a=\frac {1}{16\square }a, P_T=-\frac {1}{8\square }T \]
be formal projections which satisfy $P_c +P_a +P_T =1, P_i P_j =0, i\neq j, i,j=1,2,3 $. Possible problems with the d'Alembertian in the denominator will be disregarded for the moment.\\ 
It can be shown that the conditions $\bar D_{\dot \alpha }X=0,\dot \alpha =1,2 $, are equivalent to $ P_c X=X $ and define chiral functions (here writing $ P_c X=X $ we assume that $ X $ is such that firstly $P_c $ is applicable to it and secondly $ P_c X=X $ holds). For the coefficients chirality means

\begin{gather}
\bar \chi =\psi =n=0, v_l =i\partial_l f, \bar \lambda =-\frac{i}{2}\partial_l \varphi \sigma^l ,d=\frac{1}{4}\square f
\end{gather} 
Antichirality is characterized by $D_\alpha X=0, \alpha =1,2 $, or equivalently $P_a X=X $ or in terms of the coefficients 

\begin{gather}
\varphi =\bar \lambda =m=0, v_l =-i\partial_l f, \psi =\frac{i}{2}\sigma^l \partial _l \bar \chi ,d=\frac{1}{4}\square f
\end{gather} 
Finally, transversal functions are characterized by $D^2 X=\bar D^2 X=0$ or $P_TX=X$ or

\begin{gather}
m=n=\partial_l v^l =0, \bar \lambda =\frac{i}{2}\partial_l \varphi \sigma^l , \psi =-\frac{i}{2}\sigma^l \partial_l \bar \chi ,d=-\frac{1}{4}\square f
\end{gather}
Note the opposite signs in chiral/antichiral sectors on one side and trasversal sector on the other side. Note also that a constant supersymmetric function ($X=f,f=$constant) is at the same time chiral, antichiral and transversal. More generally, a function of the form

\begin{gather} \nonumber
X(z)=f(x)+\theta \varphi (x)+\bar \theta \bar \chi (x)\pm i\theta \sigma^l \bar \theta \partial_l f(x)
\end{gather}
such that
\[\partial_l \varphi \sigma^l =\sigma^l \partial_l \bar \chi =0,\quad \square f=0 \]
is at the same time chiral/antichiral and transversal (the particular case is the constant above which belongs to all sectors: chiral, antichiral and transversal). Such functions are concentrated on the light cone in momentum space (the constant by force too). Consequently there is no overlap between the chiral, antichiral and transversal functions on the mass hyperboloid (the massive case). \\
Note that we have payed attention to write the fermionic coefficient characterizations of the chiral, antichiral and transversal functions (which will play here the role of "test functions") in a way which is independent of assuming the components of the fermionic (spinor) coefficients to be anticommuting or commuting variables. In particular we avoid using the Pauli matrices $\bar \sigma $ and relations like $\chi \sigma^l \bar \psi =-\bar \psi \bar \sigma^l \chi $, etc. Usually in supersymmetry one assumes for concistency reasons that the components of the Weyl (Majorana) fermions are all anticommuting. This is certainly not compulsory. Alternatively we could work in the frame of a slightly more involved van der Waerden calculus in which the $\theta ,\bar \theta $-components remain Grassmann (i.e. anticommuting) whereas the components of the fermionic coefficients are $c$-numbers. Certainly the overall anticommuting convention is helpful working in the frame of the functional integral or defining Fock spaces through tensor products of supersymmetric functions (see later) but commuting fermionic components are desirable when trying to define superfields as operator-valued (super)distributions in Hilbert space as this is the standard of the rigorous quantum field theory \cite{StreW}. Working with the overall anticommuting convention we accept strange "anticommuting space-time functions". Another annoying effect is that $D_{\alpha } ,\bar D_{\dot \alpha } $ applied to $X$ of the form (2.1) is no longer of this form because the anticommutativity of the $\theta ,\bar \theta $-components with the fermionic components gets lost. Nevertheless the net results are convention independent as it should be (see later). We decide to use in this note the overall anticommuting convention and to tolerate
"anticommuting space-time functions" for the components of the fermions. If appropriate we will comment on the commuting convention too.  \\
Let $d\rho (p)$ be a (tempered) Lorentz invariant measure concentrated in the closed backward light cone in momentum space and

\begin{gather}
D^+ (x)=\frac{1}{(2\pi )^2 }\int e^{ipx}d\rho (p)
\end{gather}

\begin{gather}
K_0 (z)=\delta^2 (\theta )\delta^2 (\bar \theta )D^+ (x)\\ 
K_c (z_1 ,z_2 )=P_c K_0 (z_1 -z_2 )\\
K_a (z_1 ,z_2 )=P_a K_0 (z_1 -z_2 )\\
K_T (z_1 ,z_2 )=-P_T K_0 (z_1 -z_2 )
\end{gather}
where $ \delta^2 (\theta^2 )=\theta^2, \delta^2 (\bar \theta^2 )=\bar \theta^2 $.
Note the minus sign in front of $P_T $ which will play an important role in the sequel and the non-translation invariance of the last three kernels. Some factors of $\pi $ differ from the usual conventions.\\
For a general supersymmetric function $X$ let us decompose $ X=\begin{pmatrix}X_c \\X_a \\X_T \end{pmatrix} $ where $X_c =X_1 =P_c X ,X_a =X_2 =P_a X,X_3 =X_3 =P_T X$. We use the notation $X^T =(X_1 ,X_2 ,X_3 )$ for the transpose of $X$. We write $\mathcal{K}_0 $ for $K_0 I_3 $ where $I_3 $ is the 3x3-identity matrix. Let 

\begin{equation}\nonumber
\omega=\begin{pmatrix}1& 0 &0 \\0 &1& 0 \\0& 0 & -1 \end{pmatrix}
\end{equation}
Then we consider the inner products

\begin{equation}
<X,Y>=\int d^8z_1d^8z_2\bar X^T(z_1) \mathcal{K}_0 (z_1-z_2) Y(z_2)
\end{equation}
and

\begin{equation}
(X,Y)=<X,\omega Y>=<\omega X,Y>
\end{equation}
Let us come back to $X,Y$ in the identification $ X=\begin{pmatrix}X_c \\X_a \\X_T \end{pmatrix}, Y=\begin{pmatrix}Y_c \\Y_a \\Y_T \end{pmatrix} $. On the left hand side $X,Y$ are functions. We have in the above identification $\omega X=(P_c +P_a -P_T )X$ and after some elementary transformations we recognize $<.,.>$ and (.,.) to be

\begin{equation}
<X,Y>=\int d^8z_1d^8z_2\bar X^T (z_1)K_0 (z_1-z_2) Y(z_2)
\end{equation}
and

\begin{gather} \nonumber
(X,Y)=\int d^8z_1d^8z_2\bar X^T (z_1)(P_c +P_a -P_T )K_0 (z_1-z_2) Y(z_2)=\\ \nonumber
=\int d^8z_1d^8z_2\bar X^T (z_1)(K_c +K_a +K_T )(z_1,z_2) Y(z_2)= \\
=<X,(P_c +P_a -P_T )Y>
\end{gather}
respectively. Beside integration by parts in superspace we have used the fact that when applied to the kernel $K_0 (z_1-z_2 )$ the projection operators $P_i ,i=c,a,T $ can be freely moved from one variable to the other \cite{Sr}. One can prove \cite{C} that $<.,.>$ is indefinite (not even semidefinite) whereas (.,.) is positive definite. One assumes that the supersymmetric functions $X,Y,\ldots $ are concentrated on the support of the measure $d\rho (p)$ in order to eliminate harmless zero vectors.  \\
Some remarks are in order. First the problem with the d'Alembertian in the denominators of the formal projections. We want to look at them as projection operators in our Hilbert space. For the massive case $m\neq 0$ (in which case the measure is $ d\rho (p)=\theta (-p_0 )\delta (p^2+m^2 )$ where $\theta (p_0 ) $ is the Heaviside function) their definition as Hilbert space projection operators makes no problems (the above restriction to the support of the measure $d\rho (p)$ means the on-shell condition). The Hilbert space introduced above is well defined and decomposes into a direct sum of chiral, antichiral and transversal Hilbert subspaces. The well defined projection operators are disjoint. But in the massless case the Hilbert space scalar product is badly defined because the explicite appearence of the formal, mathematically not well-defined, projections induced by the minus sign in front of $P_T $ (although the indefinite Krein space is perfectly defined because it contains no projection operators at all; otherwise stated, they sum up to one). As a consequence the formal projections (in the massless case) cannot be generally realized any more as Hilbert space operators because of the d'Alembertian in the denominators. But there is a simple way out as follows (for more details see \cite{C}). Suppose the coefficient functions of the supersymmetric (test) functions satisfy mild restrictive conditions:

\begin{gather}
d(x)=\square D(x) \\
\bar \lambda (x)=\partial_l \Lambda (x)\sigma^l \\
\psi (x)=\sigma^l \partial_l \bar \Psi (x) \\
v_l (x)=\partial_l \rho (x) +\omega_l (x), \partial_l \omega^l =0 
\end{gather}
where $D(x), \Lambda (x), \bar \Psi (x),\rho (x), \omega (x) $ are arbitrary functions in $S$.
Just in order to have a name let us call them special supersymmetric functions. In particular the chiral, antichiral and transversal functions above are special supersymmetric.
Under these conditions the formal projections becomes well-defined. They segregate a d'Alembertian when applied to special supersymmetric functions. It follows that the non-negative inner product space deduced from the Krein structure is well defined and $P_i ,i=c,a,T $ can be realized as projection operators (they are not yet disjoint because of the sector overlap). In order to produce the desirable Hilbert space we eliminate by factorization the overlap of the chiral/antichiral and trasversal sectors which turn out to consist exactly of the zero-vectors. The chiral, antichiral and transversal sectors become disjoint. Some details on this overlap factorization apper bellow. Because we discuss in this paper the massles supervector field we place ourself from now on in the above massless situation including the massless Hilbert space $ \mathcal{H} $. In the next section we are concerned with the supersymmetric Gupta-Bleuler quantization.

\section{Gupta-Bleuler Quantization}
  
We start by defining representations of supersymmetries on supersymmetric functions by

\begin{gather}
U(g)X(z)=X(gz)
\end{gather}
where $g$ is a supersymmetric transformation. Considered in the Krein space, i.e. in the space with inner product $<.,.> $, as this is usually suggested in the literature on the subject, this representation is not unitary. We obtain a unitary representation when representing in the corresponding Hilbert space. In the massless case we have to perform the above overlap factorization too (some details appear bellow).\\
Now we come to the definition of the vector field $V(z)$ as Fock space operator. Indeed we construct the symmetric Fock space on $\mathcal{H} $ which we denote by $\mathcal{F}=Fock(\mathcal{H})$ (antisymmetric Fock spaces are reserved for ghosts). A general element of $\mathcal{F} $ will be denoted by $\Phi =(\Phi^{(0)} ,\Phi^{(1)},\ldots \Phi^{(n)}\ldots ), \Phi^{(0)}=1, \Phi^{(n)}=\Phi^{(n)}(z_1 ,z_2 \ldots ,z_n ) $. Note that working with the overall anticommuting convention discussed above, Fock spaces in supersymmetry (complying with the right statistics) are always symmetric. The above unitary representation can be extended as usually to the Fock space $\mathcal{F}$. We set for the vector field $V$ smeared with the (test) supersymmetric function $X(z)$:

\begin{gather}
V(X)=V^+(X)+V^-(X)
\end{gather}
with

\begin{gather}
(V^-(X)\Phi )^{(n)}(w_1 ,\ldots ,w_n )=\sqrt{n+1}(X(w),\Phi^{(n+1)}(w,w_1 ,\ldots ,w_n )) \\ 
(V^+(X)\Phi )^{(n)}(w_1 ,\ldots ,w_n )=\frac{1}{\sqrt n}\sum_{j=1}^n X(w_j )\Phi^{(n-1)}(w_1,\ldots ,\hat w_j
\ldots ,w_n )
\end{gather}
where $w=(p,\theta ,\bar \theta ) $, $p$ being the momentum variable conjugate to $x$, and $\hat w_j  $ means as usual omission of $w_j $. Note the appearance of the positive definite scalar product (.,.) in (3.3). In the massless case which is now under examination the positive definite scalar product, as disscussed above, requires the restriction of the supersymmetric Fock space vectors by the conditions (2.15)-(2.18) i.e. to special supersymmetric functions. This restriction (and even more) will be accomplished below. But before doing this let us remark that in the annihilation part $V^-(X)$ instead of the scalar product defined with the help of the projection $P_c +P_a -P_T $ we could use an inner product defined with the help of $\lambda_c P_c +\lambda_a P_a +\lambda_T P_T $ where $\lambda_c ,\lambda_a ,\lambda_T $ are three arbitrary real constants. In particular we could use the simplest inner product given by $P_c +P_a +P_T =1 $ (or $-P_c -P_a -P_T =-1 $) i.e. we could use the scalar product $ <.,.> $ (or $ -<.,.>) $. For $\lambda_c =\lambda_a =-\lambda ,\lambda_T =-1 $ we obtain (for the propagator) the result in \cite{WB}. In the (Feynman) gauge $\lambda =1 $ the result is particularly simple because the projections summ up to minus one but the positive definiteness of the two-point function fails. This is the result one also obtains from a computation performed on the components of the vector field. On the other hand, for $\lambda =-1$ the positivity is guaranteed by the very Hilbert-Krein structure used in this paper. In this care the two-point function is given by the kernel $K_c +K_a +K_T =(P_c +P_a -P_T )K_0 $. \\
Let us now come back to our main business and restrict our Fock space $\mathcal{F}$ to vectors $\Phi^{(n)}$ which in (each) variable $w$ are in the kernel of the operators $D^2,\bar D^2 $ (a set which as we have seen, coincides with the range of $P_T$ applied to the special functions). Formally this restriction is induced by the vanishing of operators $D^2 V^(z),\bar D^2 V^-(z) $ when applied to Fock space vectors. The conditions (2.15)-(2.18) are automatically satisfied, i.e. we are forced in the frame of special supersymmetric functions. The new (restricted) Fock space is $ \mathcal{F}'=Fock (\mathcal{H'}) $ with $\mathcal{H'}$ defined by restricting $\mathcal{H}$ to $Ker D^2 \cap Ker \bar D^2 =Im P_T $. This restriction has three positive effects: it eliminates the non-positive chiral and antichiral sectors of the Fock vectors if we start with the general inner product involving the constants $\lambda_c ,\lambda_a $, it makes applicable the projections $P_i ,i=c,a,T $ on them (because they are of the special form (2.15)-(2.18) on which $P_T $ is the identity and the d'Alembertian in the denominators of $P_c ,P_a $ is harmless), and finally it retains positivity through the minus sign of $P_T $. The operator $V(X)$ is perfectly defined albeit it does not transform $ \mathcal{F}' $ to $ \mathcal{F}' $ because $X$ is general, i.e. not special (a similar fact appears also in the classical Gupta-Bleuler method and it was traced back \cite{GW,StroW} to the fact that the vector potential is not an observable). There is still a problem with $\mathcal{F}'$, namely it contains some more zero-vectors. Indeed $Im P_T =Ker D^2 \cap Ker \bar D^2 $ has an overlap with $Ker D_{\alpha }\cup Ker \bar D_{\dot \alpha }$  and on this overlap our scalar product induced by $P_c +P_a -P_T $ vanishes. The way out is standard: we factorize these vectors from $\mathcal{F}' $ and get (by taking the closure) our final physical Fock space $\mathcal{F}_{phys} $. We realize that on $\mathcal{F}_{phys} $ all problems disappear. Summarizing: in order to obtain the physical Fock space of the supersymmetric vector field, we have to perform the factorization 

\begin{gather}
\frac{Ker D^2 \cap Ker \bar D^2 }{(Ker D_{\alpha }\cup Ker \bar D_{\dot \alpha }) \cap (Ker D^2 \cap Ker \bar D^2 ) }
\end{gather}
Using $ Ker D_{\alpha }=Im D^2 $ and $ Ker \bar D_{\dot \alpha }=Im \bar D^2 $ the above formula becomes

\begin{gather}
\frac{Ker D^2 \cap Ker \bar D^2 }{(Im D^2 \cup Im \bar D^2 )\cap (Ker D^2 \cap Ker \bar D^2 ) }
\end{gather}
It can be shown by standard Hilbert space methods that
\[Ker D^2 \cap Ker \bar D^2 =Ker D^2 /\overline {Im D^2 }=Ker \bar D^2 /\overline {Im \bar D^2 } \] where the bar means closure.
This relation will be used at the end  of this paper. \\
The reader can verity that in fact we can start the construction of the Fock space $\mathcal {F} $ on special supersymmetric functions instead of general ones. In this case the factorization needed looks as simple as

\begin{gather}
\frac{Im P_T }{(Im P_c \cup Im P_a ) \cap Im P_T }
\end{gather}
and makes use of projection operators only. An even simpler way is to modify the creation part $V^-(X)$ by introducind the projection $P_T$ acting on $X$ on the right hand sided of (3.4). In this variant the whole supersymmetric Gupta-Bleuler method would rest on the factorization of special supersymmetric functions by $Im P_c \cup Im P_a $ .\\
Note that this result and the way we have achieved it is similar but not quite identical to the classical result \cite{GW,StroW,Stro} where the kernel of the divergence contain the image of the gradient modulo wave-equation (for more comments see the next section). If we restrict the (real) massless vector field to real test functions, i.e. test functions satisfying the condition $X=\bar X $, then the matter simplifies even more. Indeed one has $Ker D^2 =Ker \bar D^2 $ and the physical Hilbert space turn out to be isomorphic to

\begin{gather}
Ker D^2 /\overline{Im D^2 }=Ker \bar D^2 /\overline {Im \bar D^2 }
\end{gather}
We want now to find gauge invariant observables in the physical Fock space. One can prove by standard methods (see for instance \cite{G}) that the supersymmetric field strengh \cite{WB} and the (massless) vector field itself are gauge invariant on $\mathcal{F}_{phys} $; the second being more important. Indeed on test functions prepared as in (3.5), (3.6) we get a gauge invariant vector field (it vanishes on the overlap of the chiral/antichiral and transversal sectors).

\section{Conclusions}

We conclude that the supersymmetric Gupta-Bleuler procedure presented in this short note deals only with elementary properties of some invariant operators (or even projection operators) which are well known in supersymmetry. Certainly in the classical case too the quantization of the electromagnetic field is intimately related to the projection operator which appears in the free Lagrangian expressed in terms of the vector potential. This projection goes at the heart of the Maxwell theory. Nevertheless in the classical Gupta-Bleuler quantization (including its rigorous version \cite{GW,StroW}) this projection does not appear explicitly. The supersymmetric situation makes explicit use of projections and as such is even simpler, in the canonical setting up, than the classical procedure. A reformulation of the present Gupta-Bleuler method using free supersymmetric ghosts (for the usual case see \cite{W}) should be possible. \\
The key and at the same time the novelty of the paper was the supersymmetric Hilbert-Krein structure. It shows its central position for purposes of the Hilbert space supersymmetric canonical quantization.\\
Acknowledgement. We thank M. Schork for discussions.

\end{document}